\documentclass[12pt,a4paper]{article}

\usepackage{style}
\usepackage{fullpage}
\usepackage[scaled=0.7]{DejaVuSansMono}

\graphicspath{{./graphics/}}

\usepackage{listings}
\definecolor{mygreen}{rgb}{0,0.5,0}
\definecolor{mygray}{rgb}{0.5,0.5,0.5}
\definecolor{mymauve}{rgb}{0.58,0,0.82}
\lstMakeShortInline[keywordstyle=\color{black},basicstyle=\fontsize{12}{12}\selectfont\ttfamily]|

\usepackage{enumitem}
\setlist[description]{leftmargin=\parindent,labelindent=\parindent}

\title{Memory Models for C/C++ Programmers}
\author{Manuel P\"oter \and Jesper Larsson Tr\"aff\\
Research Group Parallel Computing\\
Faculty of Informatics, Institute of Computer Engineering\\TU Wien\\
Favoritenstrasse 16, 1040 Wien, Austria}

\begin{document}
\maketitle
%\tableofcontents

\section{Introduction}

The memory model is the crux of the concurrency semantics of
shared-memory systems. It defines the possible values that a read
operation is allowed to return for any given set of write operations
performed by a concurrent program, thereby defining the basic
semantics of shared variables. In other words, the memory model
specifies the set of allowed outputs of a program's read and write
operations, and constrains an implementation to produce only (but at
least one) such allowed executions. The memory model may and often
does allow executions where the outcome cannot be inferred from the
order in which read and write operations occur the the program.  It is
impossible to meaningfully reason about a program or any part of the
programming language implementation without an unambiguous memory
model. The memory model defines the possible outcomes of a concurrent
programs read and write operations.  Conversely, the memory model also
defines which instruction reorderings may be permitted, either by the
processor, the memory system, or the compiler.

In this note, our programming languages will be C and C++ and programs
execute threads concurrently. Our machine consists of CPU's or cores
connected to a common, shared memory from which the cores (at the
programming language level: The threads) read and write data in some
order. We also distinguish between program (as written) and the code
(as compiled and executed by the cores).

Recent books by Michael L. Scott~\cite{Scott13} and Sorin et
al.~\cite{SorinHillWood11} provide good introductions to memory models
in both hardware and software. Overviews of the complex issues can be
found in numerous papers, for instance those by Adve and
Gharachorloo~\cite{AdveGharachorloo96}, Adve and
Boehm~\cite{Adve:2010:MMC:1787234.1787255,AdveBoehm11},
McKenney~\cite{Mckenney:MemoryOrdering}, Batty et
al.~\cite{BattyOwenSarkarSewellWeber11,BattyMemarianNienhuisPichonSewell15},
to mention a few.

A memory model can be refined to differentiate between the
\emph{programming language memory model} and the \emph{hardware memory
  model}.
\begin{itemize}
	\item Language memory model: Defines the optimizations, memory
          access re-writes and reorderings a compiler is allowed to
          perform when transforming program into code.
	\item Hardware memory model: Defines the optimizations and
          memory access reorderings a specific hardware architecture is
          allowed to perform.
\end{itemize}

These optimizations can cause memory accesses to be executed or
perceived in orders that differ from what is defined in the source
code, leading to distinguishing between the following four orderings:
\begin{description}
	\item[Source code order:] The order in which the memory operations
          are specified by the source code by the programmer.
	\item[Program order:] The order in which the memory operations
          are specified in the machine code that is executed by the
          CPU. Note that this can differ from the source code order,
          because depending on the definition of the language memory
          model, compilers are allowed to reorder instructions as
          part of the optimization process.
	\item[Execution order:] The order in which the individual
          memory-reference instructions are executed on a given
          CPU. The execution order can differ from the compiled order
          due to optimizations based on the hardware memory model of
          the specific CPU-implementation.
	\item[Perceived order:] The order in which a CPU perceives its
          and other CPUs' memory operations. The perceived order can
          differ from the execution order due to caching, interconnect
          and memory-system optimizations. Different CPUs can perceive
          the same set of memory operations as occurring in different
          orders. This is also defined by the hardware memory model.
\end{description}

The reason why these orders can be different stems from the fact that
increases in memory performance have not kept up with the rate at
which CPU instruction performance has increased. To try to hide the
fact that memory operations are increasingly expensive compared to
simple register-to-register instructions, modern CPUs are equipped
with increasingly large caches in order to reduce the overhead of
these memory accesses.

However, CPUs have become so fast that even these caches cannot keep
up with them. Therefore, caches are often partitioned into
\emph{banks} that can operate nearly independently from each
other. This allows each of the banks to run in parallel, therefore
keeping up better with the CPU. Memory usually is divided evenly among
the banks by address, e.g., even-numbered cache lines are processed by
bank $0$ while odd-numbered cache lines are processed by bank
$1$. However, this type of hardware parallelism now allows memory
operations to complete out of order.

Consider two memory write operations where the first one is processed
by bank $0$ and the second one is processed by bank $1$. Now if bank
$0$ is already busy processing an earlier request and bank $1$ is
idle, the second write would be visible to another CPU \emph{before}
the first write -- the writes would be perceived \emph{out of order}
by other CPUs. However, this kind of reordering is not limited to
write operations -- read operations can be reordered in a similar
manner.

\section{Sequential consistency}
\label{SequentialConsistency}
Sequential consistency as introduced by Lamport~\cite{Lamport79:sc} is
an (the most) intuitive memory model for reasoning about the outcome
and correctness of concurrent program, algorithm or data
structure. Many results on concurrent algorithms or data structures
either explicitly oder implicitly assume a sequentially consistent
memory model.

A natural view of the execution of a multi-threaded program is as
follows. For each step one of the threads is randomly chosen and the
next step in that thread's execution (in say, the program or the
compiled order) gets executed. This process is repeated until the
program as a whole terminates. This is effectively equivalent to
taking all the steps of all threads in (program or compiled) order,
and interleaving them in some way, resulting in a single total order
of all steps. No reordering of the thread's steps is permitted.
Therefore, whenever an object is accessed, the last value stored to
the object in this order is retrieved. An execution that can be
understood as such an interleaving is referred to as
\emph{sequentially consistent}.

Listing~\ref{lst:Dekker} shows parts of an implementation of Dekker's
mutual exclusion algorithm~\cite{Dijkstra:1965:CSP:1102034}. The steps
of the two threads can be interleaved in many ways, but since the
program order is preserved it is ensured that at least one of the load
operations sees the value of the prior store operation, i.e., the
program order, execution order and perceived order are all
identical. It is therefore not possible that both, |r1| and
|r2|, are zero.

\begin{lstlisting}[caption={Dekker's mutual exclusion algorithm}, label=lst:Dekker]
Initially: X = 0, Y = 0

Thread 1:
	X = 1
	r1 = Y
	
Thread 2:
	Y = 1
	r2 = X
\end{lstlisting}

Unfortunately ensuring sequential consistency is quite expensive and
none of todays processor architectures provide a fully sequentially
consistent memory model. While they allow to enforce sequential
consistency at certain points normal execution is not sequentially
consistent, but depends highly on the implementation of the specific
architecture.

\section{Weaker memory models}

\subsection{x86-TSO}
Even though the Intel x86 memory model is somewhat weaker than the
sequentially consistent model, it is still one of the strongest models
amongst todays modern CPU implementations. However, as Sewell et
al.~\cite{Sewell:2010:XRU:1785414.1785443} point out, for a long time
the information provided by Intel and AMD on their respective x86
architecture implementations were mostly informal, missing
concrete examples and sometimes even inconsistent with the actual
implementation.

\begin{figure}[h]
	\centering
	\includegraphics[scale=0.8]{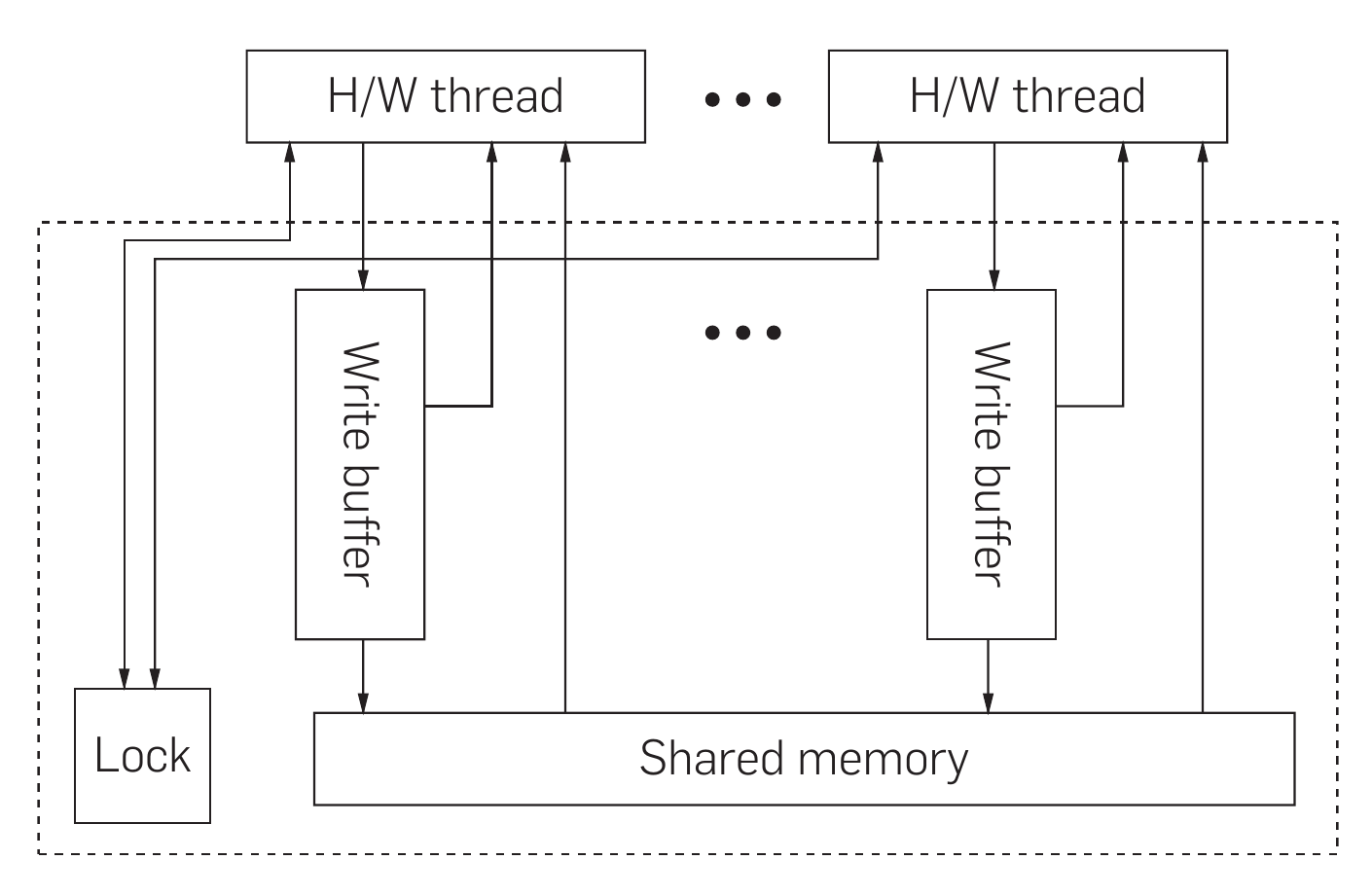}
	\caption{x86-TSO block diagram from~\cite{Sewell:2010:XRU:1785414.1785443}.}
	\label{fig:x86-TSO}
\end{figure}

Based on the available material they formally described a new memory model
called \emph{x86-TSO} (Total Store Order) which is consistent with the
concrete examples in Intel's and AMD's latest documentation available
at that time. This model is illustrated in Figure~\ref{fig:x86-TSO}.

As can be seen, the hardware threads interact with a storage subsystem
represented by the dotted box. The storage subsystem comprises a
shared memory that maps addresses to values, a global lock to indicate
when a particular hardware thread has exclusive access to memory, and
one store buffer per hardware thread. A formal definition of the
behavior of the storage subsystem can be found
in~\cite{Sewell:2010:XRU:1785414.1785443}, but the main points are:
\begin{itemize}
	\item The store buffers are FIFO and a reading thread must
          read its own most recent buffered write, if there is one, to
          that address. Otherwise reads are satisfied from shared
          memory.
	\item An |mfence| instruction flushes the store buffer of that
          thread.
	\item To execute a |lock|'d instruction\footnote{These are
          read-modify-write instructions with a \texttt{lock} prefix
          for atomicity like, e.g., \texttt{lock xadd} (atomic
          fetch-and-add) or \texttt{lock cmpxchg} (atomic
          compare-and-swap). A complete list of instructions that
          support the \texttt{lock} prefix can be found
          in~\cite[8.1.2.2]{Intel:2016:Volume3}.}, a thread must first
          acquire the global lock. At the end of the instruction, it
          flushes its store buffer and releases the lock. While the
          lock is held by one thread, no other thread can read. This
          essentially means that |lock|'d instructions enforce
          sequential consistency.
	\item A buffered write from a thread can propagate to the
          shared memory at any time except when some other thread
          holds the lock.
\end{itemize}

The model defines the perceived execution order. x86-TSO does not
permit local reordering except of reads after writes to different
addresses.

Since writes are buffered, the new value is not visible to other
threads until it has propagated to the shared memory. Therefore
Dekker's algorithm from Listing~\ref{lst:Dekker} no longer guarantees
mutual exclusion under the x86-TSO model, as it is perfectly possible
that |r1| as well as |r2| are both zero. This could be resolved by
either introducing an |mfence| instruction after the first store
operation, or by performing the store operation using a |lock xchg|
instruction.

Another memory model that is very similar to x86-TSO is the SPARC v8
TSO model~\cite{SPARCv8}.

\subsection{ARM and POWER}
\label{ARM-POWER}
The ARM and POWER architectures have considerably more relaxed memory
models, allowing a wider range of hardware optimizations. Maranget et
al.~\cite{Maranget:2012} provide a very detailed and extensive
description of both architectures and their observable behaviors.

While memory order relaxations can improve performance, power
efficiency and hardware complexity, it makes the life of a programmer,
who is implementing concurrent data structures, significantly
harder. In contrast to TSO models the following behaviors are possible
on these architectures:
\begin{enumerate}
	\item Hardware threads can perform reads and writes
          out-of-order, or even speculatively, i.e., before preceding
          conditional branches have been resolved. Any local
          reordering is allowed unless otherwise specified.
	\item The memory system does not guarantee that a write
          becomes visible to all other hardware threads at the same
          time (this behavior is called \emph{write non-atomicity}).
\end{enumerate}

Since a certain ordering of instructions is crucial already for the
simplest non-blocking data structures, these architectures provide
various memory barriers and dependency guarantees that the programmer
has to use correctly in order to enforce a desired appropriate ordering of
memory operations.

\begin{figure}[h]
	\centering
	\includegraphics[scale=0.8]{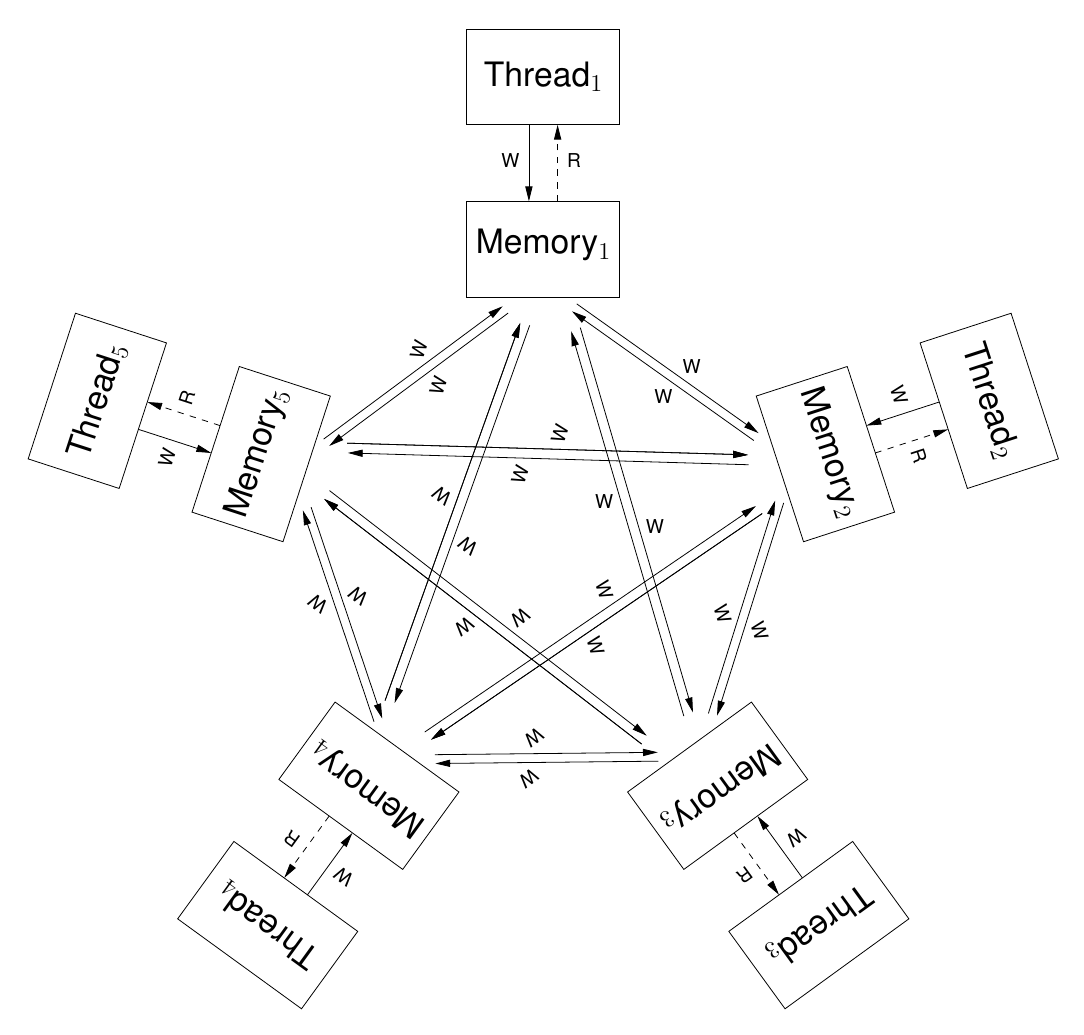}
	\caption{Storage subsystem from~\cite{Maranget:2012}.}
	\label{fig:ARM-POWER-Model}
\end{figure}

To understand the behavior of such a machine it can be helpful to
think of each hardware thread as effectively having its own copy of
memory as illustrated in Figure~\ref{fig:ARM-POWER-Model}. The
collection of all the memories and their interconnect (i.e.,
everything except the threads) is usually referred to as the
\emph{storage subsystem}. A write by one thread may propagate to other
threads in any order, and the propagations of writes to different
addresses can be interleaved arbitrarily, unless they are constrained
by barriers or cache coherence. One can also think of barriers as
propagating from the hardware thread that executed them to each of the
other threads.

The ARM |dbm| and POWER |sync| barrier (fence) instructions can be used to
enforce the following orderings between two instructions:
\begin{description}
	\item[Read/Read] the barrier ensures that they are satisfied
          and committed in program order.
	\item[Read/Write] the barrier ensures that the read is
          satisfied and committed before the write can be committed
          (and thus propagated and become visible to others).
	\item[Write/Write] the barrier ensures that the first write is
          committed and has propagated to all other threads before the
          second write is committed.
	\item[Write/Read] the barrier ensures that the write is
          committed and has propagated to all other threads before the
          read is satisfied.
\end{description}

The POWER architecture provides an additional ``lightweight sync''
instruction called |lwsync|, which is weaker and potentially faster
than |sync|. It mainly differs in how a write before the barrier is
handled relative to the second instruction:
\begin{description}
	\item[Write/Write] the barrier ensures that for any particular
          thread, the first write propagates to that thread before the
          second.
	\item[Write/Read] the barrier ensures that the write is
          committed before the read is satisfied, but the read can be
          satisfied before the write is propagated to any other
          thread.
\end{description}

In addition to barriers, these architectures provide the following
dependencies to enforce orderings:
\begin{description}
	\item[Address Dependency:] There is an address dependency from
          a read to a program-order-later read or write when the value
          read by the first instruction is used to compute the address
          of the second instruction.
	\item[Control Dependency:] There is a control dependency from a
          read to a program-order-later read/write where the value
          read by the first instruction is used to compute the
          condition of a conditional branch that is
          program-order-before the second instruction.
	\item[Data Dependency:] There is a data dependency from a read
          to a program-order-later write where the value read by the
          first instruction is used to compute the value that is
          written by the second instruction.
\end{description}

A read-to-read control dependency has little force since ARM and POWER
processors can speculatively execute past the conditional branch, thus
satisfying the second read before the first. To give a read-to-read
control dependency some force, one can add an |ISB| (ARM) or |isync|
(POWER) instruction between the conditional branch and the second read.

In contrast, a read-to-write control dependency does have some force:
the write cannot be seen by any other thread until the branch is committed,
and hence until the value of the first read is fixed.

To summarize, from one read to another, an address dependency or control
dependency with |ISB|/|isync| will prevent the second read from being
satisfied before the first one, while a plain control dependency will not.
From a read to a write, an address, control or data dependency will prevent
the write from being visible to any other thread before the value of the
read is fixed.

\section{The C++11 memory model}
\label{C++11 memory model}
On August 12\textsuperscript{th} 2011 the new C++ standard, now
commonly referred to as C++11, was approved and ratified by ISO,
replacing the previous version C++03. Since this official ISO C++
standard is not freely available we instead refer to the ``Working
Draft, Standard for Programming Language C++'' from January
2012~\cite{c++11_standard} which contains the C++11 standard plus
minor editorial changes.

This new C++ standard is the first version to define the notion of
\emph{multi-threaded executions}. The C++ Standard prior to C++11
specified program execution in terms of observable behavior, which in
turn described sequential execution on an implicitly single-threaded
abstract machine. Therefore multi-threaded C++ programs relied on
libraries for threading support, like POSIX threads, Win32, or
Boost. Unfortunately, a pure library approach in which the compiler is
designed independently of threading issues entails all sorts of
problems as discussed in the well-known paper by
Boehm~\cite{Boehm:2005:TCI:1064978.1065042}. Without a clearly defined
memory model as a common ground between the compiler, the hardware,
the threading library, and the programmer, multi-threaded C++ code is
fundamentally at odds with compiler and processor-level
optimizations~\cite{Meyers2004}. That is why with the introduction of
multi-threaded executions also a new memory model had to be defined.

The memory model defines when multiple threads may access the same
memory location, and specifies when updates by one thread become
visible to other threads. It is largely based on the work by Boehm,
Alexandrescu et
al.~\cite{Boehm:2008:FCC:1379022.1375591,Alexandrescu:2004}.

The new C++11 library defines a number of synchronization operations,
consisting of atomic operations~\cite[29, pp. 1100]{c++11_standard}
and operations on mutexes~\cite[30, pp. 1118]{c++11_standard}. These
operations play an important role in making assignments in one thread
visible to another. A synchronization operation on one or more memory
locations is either a consume, an acquire, a release, or both an
acquire and release operation.

A synchronization operations without an associated memory location is
a fence and can be either an acquire, a release, or both an acquire
and release fence. Fences are discussed in more detail in
Section~\ref{sec:Fences}.

In addition, there are relaxed atomic operations, which are not
synchronization operations, as they do not affect the visibility of
any assignment to other threads, and atomic read-modify-write
operations, which have special characteristics as explained later.

The execution model is defined on the basis of \emph{evaluations}.
There are two kinds of evaluations performed by the compiler for each
expression or subexpression (both of which are optional):
\begin{description}
	\item[Value computation:] Calculation of the value that is
          returned by the expression. This may involve determining an
          object's identity (e.g., when the expression returns a
          reference to some object) or reading the value previously
          assigned to an object (e.g., when the expression returns a
          number or some other value)
	\item[Side effect:] Access (read or write) to a |volatile| object,
          modify (write) to a (non-volatile) object, calling a
          library I/O function, or calling a function that does any of
          those operations. All such side effects are changes in the
          state of the execution environment.
\end{description}

However, conforming implementations are required to emulate (only) the
observable behavior of the abstract machine. This is often called the
``as-if'' rule, because an implementation is free to disregard any
requirement of the C++ standard as long as the result is \emph{as if}
the requirement had been obeyed, as far as can be determined from the
observable behavior of the program, i.e., it effectively \emph{allows
any and all code transformations that do not change the observable
behavior of the program}.

One of the most important aspects is the definition of a data
race~\cite[1.10.21, p. 14]{c++11_standard}:
\begin{quote}
	The execution of a program contains a data race if it contains
        two conflicting actions in different threads, at least one of
        which is not atomic, and neither \emph{happens-before} the
        other. Any such data race results in undefined behavior.
\end{quote}
Conflicting actions are defined as follows~\cite[1.10.4,
  p. 11]{c++11_standard}:
\begin{quote}
	Two expression evaluations conflict if one of them modifies a
        memory location and the other one accesses or modifies the
        same memory location.
\end{quote}
This definition implies that any program written according to the old
standard that uses some other threading libraries and shares any data
between those threads exhibits undefined behavior.  The memory
operations are ordered by means of the \emph{happens-before}
relationship that can be roughly described as follows:
\begin{quote}
	Let A and B represent operations performed by a multi-threaded
        process. If A happens-before B, then the memory effects of A
        effectively become visible to the thread performing B before B
        is performed.
\end{quote}
The \emph{happens-before} relation (denote: $\to$) is a strict partial
order and as such transitive, irreflexive and antisymmetric.
\begin{description}
	\item[transitivity:] $\forall a, b, c, \textrm{if } a \to b
          \textrm{ and } b \to c, \textrm{ then } a \to c$
	\item[irreflexivity:] $\forall a, a \not\to a$
	\item[antisymmetry:] $\forall a, b, \textrm{if } a \to b\ \textrm{ then } b \not\to a$
\end{description}

The complete formal definition specifically for C++ can be found
in~\cite[1.10, p. 11-14]{c++11_standard}.

\emph{Sequenced before} is an anti-symmetric, transitive, pair-wise
relation between instructions executed by a single thread, which
induces a partial order among those evaluations~\cite[1.9.13,
  p. 10]{c++11_standard}. Given any two evaluations $A$ and $B$, if
$A$ is sequenced before $B$, then the execution of $A$ shall precede
the execution of $B$.

One might assume that the definition of sequenced-before relation
effectively prevents the compiler from performing any instruction
reordering. But since the sequenced-before relation is only defined
between instructions executed by \emph{a single thread}, the compiler
may freely reorder instructions as long as these changes are not
observable to the executing thread (recall the ``as-if'' rule).

A \emph{happens-before} order between two operations from the same
thread (source code order) is implicitly given by the
\emph{sequenced-before} order~\cite[1.10.12, p. 13]{c++11_standard}. A
\emph{happens-before} order between two operations from different
threads (in the standard this is referred to as
\emph{inter-thread-happens-before}) must be established using atomic
operations.

Since the happens-before relation can be established between operations
in \emph{different threads}, the compiler can only reorder instructions
as long as these changes are not observable by \emph{either} thread. We
will get back to this again in Section~\ref{sec:C++11-Memory-Orders}.

\subsection{Atomic operations}
The C++11 standard library introduces a new generic class
|std::atomic<T>| that provides the following atomic operations to work
with instances of type |T|:
\begin{itemize}
	\item |load(std::memory_order order)|
	\item |store(T desired, std::memory_order order)|
	\item |exchange(T desired, std::memory_order order)|
	\item |compare_exchange_weak(T& expected, T desired, std::memory_order order)|
	\item |compare_exchange_strong(T& expected, T desired, std::memory_order order)|
\end{itemize}
For integral and pointer types it also provides the following operations:
\begin{itemize}
	\item |fetch_add(T arg, std::memory_order order)|
	\item |fetch_sub(T arg, std::memory_order order)|
\end{itemize}
And for integral types only it provides the following additional operations.
\begin{itemize}
	\item |fetch_and(T arg, std::memory_order order)|
	\item |fetch_or(T arg, std::memory_order order)|
	\item |fetch_xor(T arg, std::memory_order order)|
\end{itemize}
The |order| parameter for all operations defaults to |memory_order_seq_cst|,
but the operations can be relaxed by explicitly defining a weaker memory
order. The available memory orders and their effects are discussed in
Section~\ref{sec:C++11-Memory-Orders}.

For many of these operations the class also provides operators like
the assignment operator for |store| or post-fix increment for
|fetch_add|. These operators are a nice syntactic sugar, but they rely
on the standard memory order for all operations and do not allow
customization.

The |atomic| class can work with any type |T|, regardless of its
size. For types with a size less or equal to the size of a pointer all
the operations are usually lock-free. For other types the
implementation may fall back to a lock-based version to achieve
atomicity. The class provides the |is_lock_free| method for checking
whether the operations on the given type can be performed in a
lock-free manner.

\subsection{Memory orders}
\label{sec:C++11-Memory-Orders}
Each atomic operation takes a parameter of the type |memory_order|
which is an enumeration type with the following memory order values
(from strong to relaxed memory order):
\begin{itemize}
	\item |memory_order_seq_cst|
	\item |memory_order_acq_rel|
	\item |memory_order_release|
	\item |memory_order_acquire|
	\item |memory_order_consume|
	\item |memory_order_relaxed|
\end{itemize}

As explained in Section~\ref{SequentialConsistency}, there is a single
total order $S$ of all sequentially consistent operations. An
operation $B$ that performs a load on an object $M$ will observe the
result of the last modification $A$ of $M$ that precedes $B$ in
$S$~\cite[29.3.3, p. 1104]{c++11_standard}. From this it follows that
there is always a \emph{happens-before} relation between two
|memory_order_seq_cst| operations operating on the same object.

The orders |memory_order_consume| and |memory_order_acquire| can only
be used for operations that perform a \emph{read},
|memory_order_release| can only be used for operations that perform a
\emph{write} and |memory_order_acq_rel| can only be used for
operations that perform a \emph{read-modify-write} operation.
Although the language does not enforce these constraints some
implementations do check them at runtime\footnote{For example when the
  \lstinline|DEBUG| macro is defined the Microsoft STL implementation
  inserts code to verify these constraints at runtime.}.

A \emph{happens-before} relationship can be established by using the
following combinations of memory orders
\footnote{\lstinline|memory_order_acq_rel| is the combination of
  \lstinline|memory_order_release| and
  \lstinline|memory_order_acquire|. Wherever either of these is used it
  is also possible to use \lstinline|memory_order_acq_rel|.}:
\begin{itemize}
	\item |memory_order_seq_cst| and |memory_order_seq_cst|
	\item |memory_order_release| and |memory_order_acquire|
	\item |memory_order_release| and |memory_order_consume|
\end{itemize}

An atomic operation $A$ that performs a store-release operation on an
atomic object $M$ \emph{synchronizes with} an atomic operation $B$
that performs a load-acquire operation on $M$ and takes its value from
any side effect in the release sequence (defined below) headed by $A$.
This \emph{synchronize-with} order is compatible with the
\emph{inter-thread-happens-before} order, i.e., if $A$
\emph{synchronizes with} $B$, then $A$ \emph{inter-thread-happens-before}
$B$.

An example can be seen in
Listing~\ref{lst:acquire-release-synchronize-with}: Thread $A$ writes
two values to the two variables $x$ and $y$. In order to guarantee
that when thread $B$ sees the new value of $y$ it also sees the new
value of $x$, a \emph{happens-before} relation must be established. In
line 5 thread $A$ uses \emph{release} semantics to store the new value
of $y$ while in line 8 thread $B$ uses \emph{acquire} semantics to
load the value of $y$. If this \emph{acquire} load returns the value
stored by the \emph{release} store the two operations
\emph{synchronize with} each other, therefore establishing a
\emph{happens-before} relation. Since the store to $x$ is
\emph{sequenced before} the store to $y$ and the load of $y$ is
\emph{sequenced before} the load of $x$ it follows by transitivity
that the store to $x$ \emph{happens-before} the load of $x$.

\begin{lstlisting}[caption={Example of \emph{synchronize-with} relation with release/acquire operations.}, label=lst:acquire-release-synchronize-with]
std:.atomic<int> x, y;

// thread A
x.store(1, std::memory_order_relaxed);
y.store(2, std::memory_order_release);

// thread B
y.load(std::memory_order_acquire);
x.load(std::memory_order_relaxed);
\end{lstlisting}

As previously described, in accordance with the ``as-if'' rule the compiler
may freely reorder any instructions as long as the observable behavior for
the executing thread is consistent with the sequenced-before relation. With
regards to atomic operations, and specifically the inter-thread-happens-before
relation established via release/acquire operations, the following additional
restrictions apply:
\begin{itemize}
  \item atomic operations on the same object may never be
        reordered~\cite[1.10.19, p. 14]{c++11_standard},
  \item (non-)atomic write operations that are sequenced before a release
        operation $A$ may not be reordered after $A$,
  \item (non-)atomic load operations that are sequenced after an acquire
        operation $A$ may not be reordered before $A$.
\end{itemize}
These restrictions effectively retain the source code order of the according
instructions and thus ensure that the behavior observed by the thread that
performs the acquire operation is consistent with the happens-before relation.

The |memory_order_consume| order is based on the address dependency
concept described in Section~\ref{ARM-POWER}. As such it is not only
more complicated but also weaker than |memory_order_acquire|.
According to Hans Boehm, the current definition of
|memory_order_consume| in the standard is not
useful\footnote{\url{http://www.open-std.org/jtc1/sc22/wg21/docs/papers/2016/p0371r0.html}}. He
has therefore proposed to temporarily deprecate |memory_order_consume|
in C++17 and the proposal was accepted in the Oulu meeting in July
2016. The |memory_order_relaxed| order can never be used to establish a
\emph{happens-before} order.

All modifications to a particular atomic object occur in some
particular total order, called the \emph{modification order}. If $A$
and $B$ are modifications of an atomic object $M$ and $A$
\emph{happens-before} $B$, then $A$ precedes $B$ in the modification
order of $M$. There are separate modification orders for each atomic
object and there is no requirement that these can be combined into a
single total order for all atomic objects.

Atomic read-modify-write operations shall always read the last value
in the modification order written before the write associated with the
read-modify-write operation~\cite[29.3.12, p. 1105]{c++11_standard}.

A \emph{release sequence} is a subsequence of the modification order
of an atomic object. It is headed by a release operation $A$ and
followed by an arbitrary number of
\begin{itemize}
	\item atomic operations performed by the same thread that
          performed $A$ or
	\item atomic read-modify-write operations.
\end{itemize}
For operations performed by the same thread that performed $A$ it is
not relevant which memory order is used -- it can even use
|memory_order_relaxed|. If a thread is reading a value that is part of
a \emph{release sequence} using acquire semantics, this read
synchronizes with the release operation that is heading the sequence.
Note that there can exist several release sequences on the same object
at the same time. Suppose there are two release-CAS operation on some
atomic object $A$. Since both use release semantics, they both act as
head of their own release sequence. And since a CAS is an atomic
read-modify-write operation, the second CAS is also part of the
release sequence headed by the first CAS. So an acquire-load on $A$
that returns the value stored by the second CAS will actually
synchronize-with \emph{both} release-CAS operations.

The C++11 standard describes two different compare-and-swap operations
for atomic objects: |compare_exchange_strong| and
|compare_exchange_weak|. The difference is that\\
|compare_exchange_weak| is allowed to fail spuriously, that is, act as
if |*obj != *expected| even if they are equal, but can result in
better performance on some platforms. Both operations take two
|memory_order| parameters: The first one describes the semantics of
the read and write operations in case of success, and the second one
describes the semantics of the reload operation in the failure case.
In addition, the standard defines overloads for both operations taking
only a single |memory_order| parameter. They forward to the two
parameter version, passing the given |memory_order| as the first
argument. The second argument is also derived from the given
|memory_order| by removing any semantics that are only relevant for
write operations, i.e., |memory_order_release| is replaced with
|memory_order_relaxed| and |memory_order_acq_rel| is replaced with
|memory_order_acquire|~\cite[29.6.5.21, p. 1113]{c++11_standard}.

The C++11 standard states that ``the failure argument shall be no
stronger than the success argument''~\cite[29.6.5.20,
  p. 1113]{c++11_standard}.  But Bastien and Boehm noted that the
standard does not define the term ``stronger'' in this context, and
also questioned whether there is even a point in restricting
success/failure orderings~\cite{Bastien:2016:fail-or-succeed}. Based
on their proposal this requirement was therefore removed in C++17.

\subsection{Fences}
\label{sec:Fences}
Another synchronization operation that can be used to establish a
\emph{happens-before} relation is a \emph{fence}~\cite[29.8,
  p.\ 1116]{c++11_standard}. Just like the operations on atomic
objects, the |atomic_thread_fence| operation also takes a
|memory_order| parameter and, depending on the order, it has the
following effects:
\lstMakeShortInline[keywordstyle=\color{black},basicstyle=\fontsize{10}{10}\selectfont\ttfamily]!
\begin{itemize}
	\item has no effects, if 
          |order == memory_order_relaxed|
	\item is an \emph{acquire-fence}, if 
          |order == memory_order_acquire| or 
          |order == memory_order_consume|
	\item is a \emph{release-fence}, if 
          |order == memory_order_release|
	\item is both an \emph{acquire-fence} and a
          \emph{release-fence}, if |order == memory_order_acq_rel|
	\item is a sequentially consistent acquire- and
          release-fence, if |order == memory_order_seq_cst|
\end{itemize}
\lstDeleteShortInline!

A release fence $A$ synchronizes with an acquire fence $B$ if there
exist atomic operations $X$ and $Y$, both operating on some atomic
object $M$, such that $A$ is sequenced before $X$, $X$ modifies $M$,
$Y$ is sequenced before $B$, and $Y$ reads the value written by $X$ or
a value written by any side effect in the hypothetical release
sequence $X$ would head if it were a release operation. Alternatively
a release fence can synchronize with a load-acquire on object $M$ and
an acquire fence can synchronize with a store-release on object $M$,
given that the same sequenced before relations between the fences and
the corresponding operation are in place.

An adapted version of the previous example can be seen in
Listing~\ref{lst:acquire-release-fence-synchronize-with}. The
release-fence in line 5 is sequenced before the store operation in
line 6 and the load operation in line 9 is sequenced before the
acquire-fence in line 10. Therefore, when the load operation in line 9
returns the value written by the store operation in line 6, the
acquire-fence \emph{synchronizes with} the release-fence. From here
the happens-before relation for the operations on $x$ follows as
already described in the previous example for the release/acquire
operations in Listing~\ref{lst:acquire-release-synchronize-with}.

\begin{lstlisting}[caption={Example of \emph{synchronize-with} relation with release/acquire fences.}, label=lst:acquire-release-fence-synchronize-with]
std:.atomic<int> x, y;

// thread A
x.store(1, std::memory_order_relaxed);
std::atomic_thread_fence(std::memory_order_release);
y.store(2, std::memory_order_relaxed);

// thread B
y.load(std::memory_order_relaxed);
std::atomic_thread_fence(std::memory_order_acquire);
x.load(std::memory_order_relaxed);
\end{lstlisting}

A |memory_order_seq_cst|-fence is not only both a \emph{release} and
an \emph{acquire}-fence, but also provides some additional
properties~\cite[29.3.4-29.3.8]{c++11_standard}. They are also part of
the single total order of all sequentially consistent operations,
enforcing the following observations:
\begin{itemize}
	\item For an atomic operation $B$ that reads the value of an
          atomic object $M$, if there is a |memory_order_seq_cst|
          fence $X$ sequenced before $B$, then $B$ observes either the
          last |memory_order_seq_cst| modification of $M$ preceding
          $X$ in the total order $S$ or a later modification of $M$ in
          its modification order.
	\item For atomic operations $A$ and $B$ on an atomic object
          $M$, where $A$ modifies $M$ and $B$ takes its value, if
          there is a |memory_order_seq_cst| fence $X$ such that $A$ is
          sequenced before $X$ and $B$ follows $X$ in $S$, then $B$
          observes either the effects of $A$ or a later modification
          of $M$ in its modification order.
	\item For atomic operations $A$ and $B$ on an atomic object
          $M$, where $A$ modifies $M$ and $B$ takes its value, if
          there are |memory_order_seq_cst| fences $X$ and $Y$ such
          that $A$ is sequenced before $X$, $Y$ is sequenced before
          $B$, and $X$ precedes $Y$ in $S$, then $B$ observes either
          the effects of $A$ or a later modification of $M$ in its
          modification order.
	\item For atomic operations $A$ and $B$ on an atomic object
          $M$, if there are |memory_order_seq_cst| fences $X$ and $Y$
          such that $A$ is sequenced before $X$, $Y$ is sequenced
          before $B$, and $X$ precedes $Y$ in $S$, then $B$ occurs
          later than $A$ in the modification order of $M$.
\end{itemize}

\bibliographystyle{alpha}
\bibliography{mem-models}

\newcommand{\etalchar}[1]{$^{#1}$}
\begin{thebibliography}{BMN{\etalchar{+}}15}

\bibitem[AB10]{Adve:2010:MMC:1787234.1787255}
Sarita~V. Adve and Hans-J. Boehm.
\newblock Memory models: A case for rethinking parallel languages and hardware.
\newblock {\em Communications of the {ACM}}, 53(8):90--101, August 2010.

\bibitem[AB11]{AdveBoehm11}
Sarita~V. Adve and Hans{-}Juergen Boehm.
\newblock Memory models.
\newblock In David~A. Padua, editor, {\em Encyclopedia of Parallel Computing},
  pages 1107--1110. Springer, 2011.

\bibitem[ABH{\etalchar{+}}04]{Alexandrescu:2004}
Andrei Alexandrescu, Hans-J. Boehm, Kevlin Henney, Doug Lea, and Bill Pugh.
\newblock Memory model for multithreaded {C++}.
\newblock Document WG21/N1680=J16/04-0120; available at
  \url{http://www.open-std.org/jtc1/sc22/wg21/docs/papers/2004/n1680.pdf},
  September 2004.

\bibitem[AG96]{AdveGharachorloo96}
Sarita~V. Adve and Kourosh Gharachorloo.
\newblock Shared memory consistency models: {A} tutorial.
\newblock {\em {IEEE} Computer}, 29(12):66--76, 1996.

\bibitem[BA08]{Boehm:2008:FCC:1379022.1375591}
Hans{-}Juergen Boehm and Sarita~V. Adve.
\newblock Foundations of the {C++} concurrency memory model.
\newblock In {\em Proceedings of the {ACM} {SIGPLAN} 2008 Conference on
  Programming Language Design and Implementation {(PLDI)}}, pages 68--78, 2008.

\bibitem[BB16]{Bastien:2016:fail-or-succeed}
J.F. Bastien and Hans-J. Boehm.
\newblock Fail or succeed: there is no atomic lattice.
\newblock C++ standards committee paper,
  \url{http://www.open-std.org/jtc1/sc22/wg21/docs/papers/2016/p0418r1.html},
  August 2016.

\bibitem[BMN{\etalchar{+}}15]{BattyMemarianNienhuisPichonSewell15}
Mark Batty, Kayvan Memarian, Kyndylan Nienhuis, Jean Pichon{-}Pharabod, and
  Peter Sewell.
\newblock The problem of programming language concurrency semantics.
\newblock In {\em Programming Languages and Systems - 24th European Symposium
  on Programming ({ESOP})}, volume 9032 of {\em Lecture Notes in Computer
  Science}, pages 283--307, 2015.

\bibitem[Boe05]{Boehm:2005:TCI:1064978.1065042}
Hans{-}Juergen Boehm.
\newblock Threads cannot be implemented as a library.
\newblock In {\em Proceedings of the {ACM} {SIGPLAN} 2005 Conference on
  Programming Language Design and Implementation {(PLDI)}}, pages 261--268,
  2005.

\bibitem[BOS{\etalchar{+}}11]{BattyOwenSarkarSewellWeber11}
Mark Batty, Scott Owens, Susmit Sarkar, Peter Sewell, and Tjark Weber.
\newblock Mathematizing {C++} concurrency.
\newblock In {\em Proceedings of the 38th {ACM} {SIGPLAN-SIGACT} Symposium on
  Principles of Programming Languages ({POPL})}, pages 55--66, 2011.

\bibitem[CSC12]{c++11_standard}
Stefanus Du~Toit C++ Standards~Committee.
\newblock Working draft, standard for programming language c++.
\newblock C++ standards committee paper,
  \url{http://www.open-std.org/jtc1/sc22/wg21/docs/papers/2012/n3337.pdf},
  January 2012.

\bibitem[Dij65]{Dijkstra:1965:CSP:1102034}
Edsger~Wybe Dijkstra.
\newblock Cooperating sequential processes.
\newblock Technical Report EWD-123, X, 1965.

\bibitem[{Int}16]{Intel:2016:Volume3}
{Intel Corporation}.
\newblock {\em Intel® 64 and IA-32 Architectures Software Developer’s Manual
  Volume 3 (3A, 3B, 3C \& 3D): System Programming Guide}, September 2016.

\bibitem[Lam79]{Lamport79:sc}
Leslie Lamport.
\newblock How to make a multiprocessor computer that correctly executes
  multiprocess procegrams.
\newblock {\em {IEEE} Computer}, 28(9):690--691, 1979.

\bibitem[MA04]{Meyers2004}
Scott Meyers and Andrei Alexandrescu.
\newblock {C++} and the perils of double-checked locking.
\newblock {\em Doctor Dobb’s Journal}, Jul 2004.

\bibitem[McK05]{Mckenney:MemoryOrdering}
Paul~E. McKenney.
\newblock Memory ordering in modern microprocessors.
\newblock {\em Linux Journal}, 30:52--57, 2005.

\bibitem[MSS12]{Maranget:2012}
Luc Maranget, Susmit Sarkar, and Peter Sewell.
\newblock A tutorial introduction to the {ARM} and {POWER} relaxed memory
  models.
\newblock Technical report,
  \url{https://www.cl.cam.ac.uk/~pes20/ppc-supplemental/test7.pdf}, October
  2012.

\bibitem[Sco13]{Scott13}
Michael~L. Scott.
\newblock {\em Shared-Memory Synnchronization}.
\newblock Synthesis Lectures on Computer Architecture. Morgan \& Claypool
  Publisher, 2013.

\bibitem[SHW11]{SorinHillWood11}
Daniel~J. Sorin, Mark~D. Hill, and David~A. Wood.
\newblock {\em A primer on Memory Consistency and Cache Coherence}.
\newblock Synthesis Lectures on Computer Architecture. Morgan \& Claypool
  Publisher, 2011.

\bibitem[{SPA}92]{SPARCv8}
{SPARC International Inc.}
\newblock {\em The SPARC Architecture Manual Version 8 Revision SAV080SI9308},
  1992.

\bibitem[SSO{\etalchar{+}}10]{Sewell:2010:XRU:1785414.1785443}
Peter Sewell, Susmit Sarkar, Scott Owens, Francesco~Zappa Nardelli, and
  Magnus~O. Myreen.
\newblock {X86-TSO}: A rigorous and usable programmer's model for x86
  multiprocessors.
\newblock {\em Communications of the {ACM}}, 53(7):89--97, 2010.

\end{thebibliography}

\end{document}